\documentclass[prd,nofootinbib,english]{revtex4}

\usepackage{graphicx,float}
\usepackage{amsmath,amssymb,amsfonts}
\usepackage{epsfig,color}
\usepackage[thinlines]{easytable}
\usepackage{pdfpages}
\usepackage{array}
\usepackage{cancel}
\usepackage{mathtools}
\usepackage{accents}
\usepackage{subfigure}
\usepackage{enumitem}
\usepackage[dvipsnames]{xcolor}
\usepackage{refstyle}
 \usepackage{hyperref}
\hypersetup{
    colorlinks=true,
    linkcolor=blue,
    filecolor=magenta,      
   citecolor=blue
}

\begin{document}

\title{Covariant formulation of $f(Q)$ theory}

\author{Dehao Zhao}
\email{dhzhao@mail.ustc.edu.cn}
\affiliation{Interdisciplinary Center for Theoretical Study,
	University of Science and Technology of China, Hefei, Anhui 230026, China}
\affiliation{Peng Huanwu Center for Fundamental Theory, Hefei, Anhui 230026, China}

\begin{abstract}
In Symmetric Teleparallel General Relativity, gravity is attributed to the non-metricity. The so-called "coincident gauge" is usually taken in this theory so that the affine connection vanishes and the metric is the only fundamental variable. This gauge choice was kept in many studies on the extensions of Symmetric Teleparallel General Relativity, such as the so-called $f(Q)$ theory. In this paper, we point out that sometimes this gauge choice conflicts with the coordinate system we selected based on symmetry. To circumvent this problem, we formulate the $f(Q)$ theory in a covariant way with which we can find suitable non-vanishing affine connection for a given metric. We also apply this method to two important cases: the static spherically symmetric spacetime and the homogeneous and isotropic expanding universe.
\end{abstract}

\maketitle

\section{Introduction}

Although Einstein's General Relativity (GR) has achieved great success, the cosmological observations posed new challenges such as the dark energy and dark matter problems. There are usually two approaches to attack these problems, the first one is to modify the matter sector by adding some additional `dark' components to the energy budget of the universe, and the second one is to modify GR. In the second approach, while most works adopt standard curvature-based formulation, e.g., extending the Einstein-Hilbert action to a general $f(R)$ action, there exist other ways based on those theories equivalent to GR, such as Teleparallel Equivalent of General Relativity (TEGR) \cite{A. Einstein, Sauer:2004hj}, and Symmetric Teleparallel General Relativity (STGR) \cite{Nester:1998mp, BeltranJimenez:2017tkd,BeltranJimenez:2018vdo}. 

TEGR is a gravity theory equivalent to GR, which attributes the gravity to the torsion instead of curvature of the spacetime. It is formulated in flat spacetime with torsion, and in general the basic variables are the tetrad and spin connection if we use the tetrad language, where the spin connection is constrained by the vanishing of the curvature tensor and the non-metricity tensor. Theses two constrains allows us to choose the so-called Weitzenb\"{o}ck connection:  i.e., all the components of the spin connections vanish, so that the tetrad components are the only fundamental variables. This is considered as a gauge choice in TEGR and brings no change in physics, since any other choices compatible with the constraint of teleparallelism will lead to a same action up to a surface term.  However, in modified TEGR such as $f(T)$ theory\footnote{See Ref. \cite{Cai:2015emx} for a review}, the well-known `good and bad tetrads' problem \cite{Tamanini:2012h good and bad,Ferraro:2011us} appears, and the theory will violate the local Lorentz invariance if we still choose the Weitzenb\"{o}ck connections and  only consider the tetrad as basic variable. This means any two tetrads that differ by a local Lorentz transformation are not physically equivalent and some of tetrads will not satisfy the equations of motion. Then we can see that  if we use Weitzenb\"{o}ck connection in those theories, how to choose suitable `good tetrads' becomes a question. There are some works have tried to solve this problem \cite{Tamanini:2012h good and bad,Ferraro:2011us}, but the processes are complicated and not universal. Hence, in some recent works the authors began to abandon the Weitzenb\"{o}ck connection, and solved modified TEGR theory in a covariant way \cite{Krssak:2015oua,Krssak:2018ywd,Hohmann:2018rwf,Hohmann:2019fvf,Hohmann:2020zre,Hohmann:2019nat} in which both tetrad and spin connection are treated as fundamental variables under the teleparallelism constraint. Therefore the problem of searching for a `good tetrad' is replaced by looking for a suitable connection for any given tetrad to satisfy the equations of motion. This is relatively easier than the `good and bad tetrads'  problem.  

STGR is a formulation equivalent to GR which shares many similar properties with TEGR but is less considered. In STGR  theory, the curvature and torsion are set to zero, and the gravity is attributed to the non-metricity tensor $Q_{\alpha\mu\nu}$, which depends the metric and affine connection. Similar to TEGR, here under the teleparallelism and torsion-free constraints, we can always choose the so-called `coincident gauge' \cite{BeltranJimenez:2017tkd} in which the affine connection vanishes. Again, in STGR this is indeed a gauge condition because other choices only contribute a surface term in the action. But in modified STGR theories such as the so-called $f(Q)$ theory, if we still choose the coincident gauge \cite{Lu:2019hra,Jimenez:2019ovq,Harko:2018gxr,Frusciante:2021sio} and take metric as the only fundamental variable, the evolution of metric will be different in different coordinate systems. We will point out in this paper that the coordinate system we selected based on the symmetries of spacetime may be not compatible with the coincident gauge in the sense of equations of motion. So just like the `good and bad tetrads' problem in modified TEGR, we can call this problem as the `good and bad coordinate systems' problem in modified STGR theories. Also like in modified TEGR theories, in this paper, we do not look for which coordinate systems are compatible with coincident gauge, but we will use $f(Q)$ theory as an example and present a method to search for suitable affine connections for arbitrary coordinate systems in modified STGR theories.  

The outline of this paper is as follows. In Section \ref{section STGR}, we will give a brief review of STGR. In Section \ref{section fQ}, we will briefly introduce the simple modified STGR theory, $f(Q)$ theory, and show how the inconsistency appears when naively taking the coincident gauge. We will propose our method to circumvent this difficulty. In Section \ref{section solution}, we will apply our covariant method to two important cases: the spherically symmetric spacetime and the homogeneous and isotropic expanding universe. Section \ref{summery} is the conclusion.

\section{Symmetric Teleparallel General Relativity}\label{section STGR}

In this section, we will give a brief review of Symmetric Teleparallel General Relativity ( STGR ) theory. Let us start from the general metric-affine theory with a manifold  $(M, g_{\mu \nu}, {\Gamma^{\lambda}}_{\mu \nu})$, where  $g_{\mu\nu}$ is metric tensor with signature $(-1, +1, +1, +1)$ and ${\Gamma^{\lambda}}_{\mu \nu}$ (or $\nabla$) represents an arbitrary affine connection. The torsion and non-metricity tensors are defined as
\begin{eqnarray}
	{T^{\lambda}}_{\mu\nu} :={\Gamma^{\lambda}}_{\mu \nu} - {\Gamma^{\lambda}}_{\nu\mu}, \quad	Q_{\alpha \mu \nu} :=\nabla_{\alpha} g_{\mu \nu} = \partial_{\alpha} g_{\mu\nu} - {\Gamma^{\lambda}}_{\alpha\mu}g_{\lambda\nu} - {\Gamma^{\lambda}}_{\alpha\nu}g_{\mu\lambda},
\end{eqnarray}
then the affine connection can be represented as
\begin{eqnarray} \label{Gamma=Gamma+S}
	{\Gamma^{\lambda}}_{\mu \nu}  =  {\mathring{\Gamma}^{\lambda}}{}_{\mu \nu} + {S^{\lambda}}_{\mu\nu},
\end{eqnarray}
where ${\mathring{\Gamma}^{\lambda}}{}_{\mu \nu}$ is the usual Levi-Civita connection
\begin{eqnarray}
	{\mathring{\Gamma}^{\lambda}}{}_{\mu \nu} = \frac{1}{2} g^{\lambda\rho}( \partial_{\mu}g_{\rho\nu} + \partial_{\nu}g_{\mu\rho} - \partial_{\rho}g_{\mu\nu} ),
\end{eqnarray}
and 
\begin{eqnarray}
	S_{\lambda\mu\nu}= -\frac{1}{2}( T_{\mu\nu\lambda} +T_{\nu\mu\lambda} - T_{\lambda\mu\nu} ) -\frac{1}{2}( Q_{\mu\nu\lambda} +Q_{\nu\mu\lambda} - Q_{\lambda\mu\nu} )
\end{eqnarray}
is called the distortion tensor. We use $\mathring{\nabla}$ as the covariant derivative operator corresponding to the Levi-Civita connection ${\mathring{\Gamma}^{\lambda}}{}_{\mu \nu}$ and also from now on, all quantities which denoted by a ring over the heads are associated with the Levi-Civita connection, unless otherwise specified. The relation between curvatures corresponding to ${\Gamma^{\lambda}}_{\mu \nu}$  and  ${\mathring{\Gamma}^{\lambda}}{}_{\mu \nu}$ is
\begin{eqnarray} \label{curvature relation}
	{R_{\mu\nu\rho}}^{\sigma }  = {\mathring{R}_{\mu\nu\rho}}{}^{\sigma } - 2 \mathring{\nabla}_{[\mu}{S^{\sigma}}_{\nu]\rho}+2{S^{\lambda}}_{[\mu|\rho|}{S^{\sigma}}_{\nu]\lambda},
\end{eqnarray}
and the scalar curvature relation
\begin{eqnarray}\label{R=R+}
	R =g^{\mu\rho} {R_{\mu\sigma\rho}}^{\sigma }= \mathring{R} - 2\mathring{\nabla}_{[\mu}{{S^{\sigma}}_{\sigma]}}^{\mu} + 2{{S^{\lambda}}_{[\mu}}^{\mu}{S^\sigma}_{\sigma]\lambda}.
\end{eqnarray}
In STGR theory, the affine connection is constrained by the curvature-free and torsion-free conditions. The curvature-free condition requires Riemann tensor ${R_{\mu\nu\rho}}^{\sigma}(\Gamma)$ is zero, and Eq.(\ref{R=R+}) becomes 
\begin{eqnarray}\label{T=}
	-2{{S^{\lambda}}_{[\mu}}^{\mu}{S^\sigma}_{\sigma]\lambda}=\mathring{R} - 2\mathring{\nabla}_{[\mu}{{S^{\sigma}}_{\sigma]}}^{\mu}.
\end{eqnarray}
Since the Riemann tensor vanishes, the parallel transport defined by the covariant derivative $\nabla$ and its associated affine connection ${\Gamma^{\lambda}}_{\mu \nu}$ is independent of the path, which is the reason for the terminology -- 'teleparallel'. Besides the condition of zero curvature, this theory further poses a torsionless constraint on the connection, ${T^{\lambda}}_{\mu\nu}=0$, so that in symmetric teleparallel gravity theory the gravitation is totally attributed to the non-metricity. Since torsion tensor vanishes, affine connection is symmetric in its lower indices and which is the reason for the terminology -- 'symmetric'. Finally, using Eq.(\ref{T=}), we can write down a GR equivalent action which is entirely constructed from the non-metricity tensor
\begin{eqnarray}\label{action-g}
	S_{g} = \frac{1}{2}\int d^{4}x \sqrt{-g} Q = \frac{1}{2}\int d^{4}x \sqrt{-g}  Q_{\alpha\mu\nu}P^{\alpha\mu\nu}~, 
\end{eqnarray}
which is the gravitational part of STGR action, here we have set $8\pi G=1$ and the tensor $P^{\alpha\mu\nu}$ is related to scalar $Q$ as
\begin{eqnarray}
	P^{\alpha\mu\nu} = -\frac{1}{4}Q^{\alpha\mu\nu} + \frac{1}{2}Q^{(\mu\nu)\alpha}+\frac{1}{4}(Q^{\alpha} - \tilde{Q}^{\alpha})g^{\mu\nu} -\frac{1}{4} g^{\alpha(\mu}Q^{\nu )}~,
\end{eqnarray}
where the vectors $Q_{\alpha}, \tilde{Q}_{\alpha}$ are two different traces of the non-metricity tensor 
\begin{eqnarray}
	Q_{\alpha} = g^{\sigma\lambda}Q_{\alpha\sigma\lambda}, \quad  \tilde{Q}_{\alpha} =  g^{\sigma\lambda}Q_{\sigma\alpha\lambda}~.
\end{eqnarray}
Using the relation of Eq.(\ref{T=}), one can rewrite this action (\ref{action-g}) as
\begin{equation}\label{action-g2}
	S_g= \frac{1}{2}\int d^{4}x   \sqrt{-g}\left[  \mathring{R}+\mathring{\nabla}_{\mu}(Q^{\mu} - \tilde{Q}^{\mu}) \right] .
\end{equation}
where the first term is the Einstein-Hilbert action of GR and second term is a surface term which can be dropped. Therefore STGR is equivalent to GR

As mentioned above, in the process of constructing the action of STGR, we used the constraints ${R_{\mu\nu\rho}}^{\sigma}(\Gamma) = 0$ and ${T^{\lambda}}_{\mu\nu}=0$. As pointed out in Ref.\cite{Bao:2000}, these conditions allow people to choose a coordinate system $\left\{ y^{\mu}\right\}$ such that the affine connection ${\Gamma^{\lambda}}_{\mu \nu} (y^{\mu})$ vanishes,  which is the so-called coincident gauge \cite{BeltranJimenez:2017tkd}. Then in any other coordinate system $\left\{ x^{\mu}\right\}$, the affine connection has the following form,
\begin{eqnarray} \label{gamma=}
	{\Gamma^{\lambda}}_{\mu \nu} (x^{\mu}) = \frac{\partial x^{\lambda}}{\partial y^{\beta}} \partial_{\mu}\partial_{\nu} y^{\beta}~.
\end{eqnarray}
Since there is a coordinate system $\left\{ y^{\mu}\right\}$ in which the affine connection vanishes, we can always think that we are working in this special coordinate system and the only fundamental variable is metric. Choosing a special coordinate system usually means the breaking of the diffeomorphism symmetry.  But this does not happen for STGR theory. Because in the beginning,  we don't know in which coordinate system, the affine connection is zero, so we can take an arbitrary coordinate system A and say that in A affine connection is zero. Also in the beginning, we can think in another coordinate system B, affine connection is zero. It is easy to found that the differences of action between this two case are only a total derivative. We know that the surface term has no effect on the equations of motion (EOMs). So it means that in the beginning, no matter which coordinate system we think affine connection is zero, the evolution of metric is unchanged. This is the meaning of diffeomorphism symmetry. Varying the action (\ref{action-g}) with respect to the metric, we get the EOMs of STGR
\begin{eqnarray}\label{STGR eoms}
	\frac{2}{\sqrt{-g}} \partial_{\lambda} \left(\sqrt{-g} P^{\lambda\mu\nu}\right) -\frac{1}{2} Qg^{\mu\nu} +\left(Q^{\mu\rho\sigma}{P^{\nu}}_{\rho\sigma} + 2Q^{\rho\sigma\mu}{P_{\rho\sigma}}^{\nu}\right)	= \tau^{\mu\nu},
\end{eqnarray}
where $\tau^{\mu\nu}$ is energy-momentum tensor which is defined as
\begin{eqnarray}
	\tau^{\mu\nu} = \frac{2}{\sqrt{-g}} \frac{\delta S_{m}}{\delta g_{\mu\nu}},
\end{eqnarray}
where $S_m$ is matter's action. In next section, we can see that Eq.(\ref{STGR eoms}) and Einstein equation are equivalent.

\section{Covariant $f(Q)$ theory} \label{section fQ}

From the discussions in the previous section, we have known that original STGR theory owns the same dynamics as GR and no matter which connections we choose, the evolution of metric is unchanged. But in modified STGR theory, such as $f(Q)$ theory where the $Q$ scalar in the action (\ref{action-g}) is replaced by its general function, the situation is different. For that case, the term of action which depends on the affine connection is no longer a total derivative. So different affine connections leads to different evolution of metric, then leads to different solutions. Therefore in modified STGR theories, how to choose a better and meaningful affine connection becomes an important question and it is also the central content of this paper. In this section we lay out the suggestions to circumvent this question and first we briefly introduce some basic aspects about $f(Q)$ theory. 

\subsection{Action and equations of motion} \label{action and eoms}

The action of $f(Q)$ theory is
\begin{eqnarray}
	S= \frac{1}{2}\int d^{4}x \sqrt{-g} f(Q) + S_{m} ~.
\end{eqnarray}
This theory can be considered as a metric-affine theory which takes $g_{\mu\nu}$ and ${\Gamma^{\lambda}}_{\mu \nu}$ both as fundamental variables. So to get the EOMs of this theory, we should do variations with respect to them separately. Meanwhile, because this theory should satisfies the curvature-free and torsion-free conditions, the affine connection can be always written as Eq.(\ref{gamma=}), which means that affine connection in this theory can be totally determined by four arbitrary functions $y^{\mu}$. Thus to get the equations of motion, we can consider $g_{\mu\nu}$ and $y^{\mu}$ as fundamental variables and variate the action with respect to them. If we consider the matter's action only relies on metric and matter fields, the metric's EOMs of $f(Q)$ are
\begin{eqnarray}
	E_{\mu\nu} \equiv\frac{2}{\sqrt{-g}} \nabla_{\lambda}(\sqrt{-g} f_{Q} {P^{\lambda}}_{\mu\nu} ) - \frac{1}{2}fg_{\mu\nu} +f_{Q}(P_{\nu\rho\sigma}{Q_{\mu}}^{\rho\sigma} - 2P_{\rho\sigma\mu}{Q^{\rho\sigma}}_{\nu})  =\tau_{\mu\nu}   \label{metric f(Q) eq} , 	
\end{eqnarray}
where $f_{Q} = \partial f/\partial Q$ and for later convenience, we denoted 
the EOMs of metric as $E_{\mu\nu}$. Since $f(Q)$ theory is a metric-affine theory, the solutions of this theory should contain both metric and STGR covariant derivative which satisfies the curvature-free and torsion-free conditions. Since in GR we only need metric, we think this fact reflects that in modified STGR theories one has opportunities to find more solutions than GR. 

The process of getting the variation with respect to metric is straightforward, but the variation with respect to $y^{\mu}$ need some tricks. To get the EOMs of $y^{\mu}$, let us first consider a diffeomorphism transformation which corresponds to a general infinitesimal coordinate transformation, $y^{\mu} \rightarrow y^{\mu}+ \zeta(y)$. After some calculations, it can be found that under this diffeomorphism transformation, the change of affine connection is 
\begin{eqnarray}
	\delta_{\zeta} {\Gamma^{\lambda}}_{\mu\nu}=\mathcal{L}_{\zeta} {\Gamma^{\lambda}}_{\mu\nu} = - \nabla_{\mu}\nabla_{\nu} \zeta^{\lambda}.
\end{eqnarray} 
Now, we analyze the meanings of this result. If in coordinate system $\left\{y^{\mu}\right\}$, the affine connection ${\Gamma^{\lambda}}_{\mu\nu} $ corresponding to $\nabla$ vanishes, then a diffeomorphism transformation will lead to a new coordinate system $\tilde{y}^{\mu}=y^{\mu}+ \zeta(y)$ and a new STGR covariant derivative $\tilde{\nabla}$ which affine connection ${\tilde{\Gamma}^{\lambda}}{}_{\mu\nu}$ vanishes in the new coordinate system $\left\{\tilde{y}^{\mu}\right\}$. This means that the variation of connection with respect to $y^{\mu}$ is the same as the change of connection under a diffeomorphism transformation, that is,
\begin{eqnarray}\label{variation equal transformation}
	\delta_{y}{\Gamma^{\lambda}}_{\mu\nu}=\delta_{\zeta} {\Gamma^{\lambda}}_{\mu\nu}= - \nabla_{\mu}\nabla_{\nu} \zeta^{\lambda}.
\end{eqnarray} 
then we can get the final EOMs of $y^{\mu}$,
\begin{eqnarray}\label{connection equations}
	\nabla_{\nu}\nabla_{\mu}( \sqrt{-g} f_{Q} {P^{\mu\nu}}_{\lambda} ) = 0 \label{zeta equation}.
\end{eqnarray}             

Now, we want to point out an important result that the EOMs of $y^{\mu}$ can be derived from the EOMs of metric and matter. The proof is as follows. If we consider the matter is a scalar field $\phi$, that is $S_{m}=S_{m}(g_{\mu\nu},\phi)$, then do a diffeomorphism transformation,  $x^{\mu} \rightarrow x^{\mu}+ \zeta(x)$, the change of action is zero,
\begin{eqnarray}
	\delta_{\zeta}S = \frac{\delta S}{\delta g}|_{\Gamma, \phi} \delta_{\zeta}g + \frac{\delta S}{\delta \phi}|_{g,\Gamma}\delta_{\zeta}\phi+\frac{\delta S}{\delta \Gamma}|_{g,\phi} \delta_{\zeta}\Gamma=0.
\end{eqnarray}
Therefore if the EOMs of matter and metric are satisfied, i.e. $\frac{\delta S}{\delta \phi}|_{g,\Gamma}=0$ and $\frac{\delta S}{\delta g}|_{\Gamma, \phi}=0$, then $\frac{\delta S}{\delta \Gamma}|_{g,\phi} \delta_{\zeta}\Gamma=0$ is satisfied automatically. Meanwhile because the variation of connection with respect to $y^{\mu}$ is the same as the change of connection under a diffeomorphism transformation i.e. Eq.(\ref{variation equal transformation}), then $\frac{\delta S}{\delta \Gamma}|_{g,\phi} \delta_{y}\Gamma=0$ is satisfied automatically, which is the EOMs of $y^{\mu}$. So in the following contents, we only consider the metric's EOMs.

As proved in \ref{prof of nice formula}, the EOMs of metric i.e. Eq.(\ref{metric f(Q) eq}) can be rewritten as
\begin{eqnarray} \label{new form EOMs}
	f_{Q} \mathring{G}_{\mu\nu} +\frac{1}{2} g_{\mu\nu} (f_{Q}Q - f) +2f_{QQ}(\partial_{\lambda}Q) {P^{\lambda}}_{\mu\nu} = \tau_{\mu\nu},
\end{eqnarray}
where $\mathring{G}_{\mu\nu} = \mathring{R}_{\mu\nu} -1/2\mathring{R}g_{\mu\nu}$ is the Einstein tensor corresponding to the Levi-Civita connection. From this formula, we can better understand the problem of diffeomorphism invariance. For STGR, $f(Q)=Q$, the left hand of Eq.(\ref{new form EOMs}) is just the Einstein tensor which only relies on the metric, so just as we have discussed in previous section, no matter which affine connection we choose, it has no influence on the evolution of metric. But in more general cases in which $f(Q)$ is not the linear function of $Q$, the affine connection will certainly enter the dynamics of the metric.

\subsection{Try to find solutions}

Since we have obtained the equations of motion in covariant $f(Q)$ theory, the next question will be how to find solutions. The covariant $f(Q)$ theory is diffeomorphism invariant where metric and connection both are basic variables. If one find a solution $\{ g_{\mu\nu}, {\Gamma^{\lambda}}_{\mu \nu} \}$ for this theory, then after a coordinate transformation $x\rightarrow \tilde{x}$, the result
\begin{eqnarray}
	\tilde{g}_{\mu\nu}=\frac{\partial x^{\alpha}}{\partial\tilde{x}^{\mu}} \frac{\partial x^{\beta}}{\partial \tilde{x}^{\nu}}g_{\alpha\beta}, \quad
	{\tilde{\Gamma}^{\lambda}}{}_{\mu\nu}= \frac{\partial^2 x^{\beta}}{\partial\tilde{x}^{\mu} \partial\tilde{x}^{\nu}} \frac{\partial \tilde{x}^{\lambda}}{\partial x^{\beta}} + {\Gamma^{\beta}}_{\alpha\sigma} \frac{\partial x^{\alpha}}{\partial\tilde{x}^{\mu}}\frac{\partial x^{\sigma}}{\partial\tilde{x}^{\nu}}\frac{\partial \tilde{x}^{\lambda}}{\partial x^{\beta}}
\end{eqnarray} 
is also a solution. GR is well-studied, the metric ansatz of many special cases have already obtained. Therefore in the covariant $f(Q)$ theory, problem will be how to choose a suitable affine connection for a given ansatz of metric. 

A straightforward idea is that since the curvature-free and torsion-free conditions always allow us to choose a coordinate system $\left\{ y^{\mu}\right\}$ in which  ${\Gamma^{\lambda}}_{\mu \nu} (y^{\mu})= 0$, we can just choose a coordinate system (such as in spherically symmetric spacetime, we choose the spherically symmetric coordinate system ), and say that in this coordinate system affine connection vanishes. But from the following examples, we found this simple idea filed.

\subsubsection{ Static spherically symmetric spacetime}

Static spherically symmetric spacetime is an important and frequently studied case for theories of gravity, because it is the basis for our understandings of many astronomical phenomena. In static spherically symmetric spacetime, a natural choice of coordinate system is the spherically symmetric coordinate system. The metric ansatz we choose in this coordinate system is   
\begin{eqnarray} \label{schwartz metric}
	ds^2 = -e ^{A(r)} dt^2 + e^{B(r)}dr^2 + r^2( d\theta^2 + \sin^2\theta d\phi^2).
\end{eqnarray}
If we assume the affine connection is zero in this coordinate system and further require $f(Q)$ theory has vacuum solutions, this is $E_{\mu\nu}=0$, then one can read out the off-diagonal component of Eq.(\ref{metric f(Q) eq}) are
\begin{eqnarray}\label{off-diagonal sphere}
	E_{r\theta }=E_{\theta r} = -\frac{1}{2}f_{QQ}  \frac{\partial Q}{\partial r} \cot\theta =0~.
\end{eqnarray}
where $Q=2e^{-B} (1+r A'(r))/r^{2}$, and $A'(r)=d A(r)/dr$. Those equations together with the diagonal components of Eq.(\ref{metric f(Q) eq}) consequently gives the result $f_{QQ}=0$. This result means that if at the beginning, choosing $f(Q)=Q^2$ will lead to inconsistent EOMs. So for those theories where $f(Q)$ are not linear functions of $Q$, metric Eq.(\ref{schwartz metric}) with affine connection ${\Gamma^{\lambda}}_{\mu\nu} =0$ is not a solution of EOMs. It does not mean that $f(Q)$ theory does not contain static spherically symmetric vacuum solutions, but means that the spherically symmetric coordinate system is not compatible with the coincident gauge. 

One may wonder that when we fix coincident gauge, choosing different coordinate systems leads to different evolution of metric.  So if we want to get spherically symmetric solutions, we should assume a more general form of the static spherically symmetric metric
\begin{eqnarray} \label{schartz C}
	ds^2 = -e ^{A(r)} dt^2 + e^{B(r)}dr^2 + C(r)^2( d\theta^2 + \sin^2\theta d\phi^2).
\end{eqnarray}
In normal circumstances, indeed, we should do that. But for the special $f(Q)$ case, it seems that the theory still features a  reparameterization symmetry of radius $r$. This fact can be seen from EOMs of $f(Q)$ theory. We choose coincident gauge and consider metric as the only fundamental variable. We denote the EOMs that obtained from metric Eq.(\ref{schartz C}) as $E_{\mu\nu}$ and denote the EOMs obtained from metric
\begin{eqnarray} \label{schartz C2}
	ds^2 = -e ^{A(r)} dt^2 + e^{M(r)}dC^2 + C(r)^2( d\theta^2 + \sin^2\theta d\phi^2)
\end{eqnarray}
as $\tilde{E}_{\mu\nu}$, where $e^{M(r)}=e^{B(r)}/C'(r)^{2}$, then one can find the difference between $E_{\mu\nu}$ and $\tilde{E}_{\mu\nu}$ is just a coordinate transformation $\tilde{E}_{\mu\nu}=(\partial x^{\alpha}/\partial\tilde{x}^{\mu})( \partial x^{\beta}/\partial \tilde{x}^{\nu}) E_{\alpha\beta}$ and this  coordinate transformation is a reparameterization of radius $r\rightarrow C(r)$. So the EOMs derived from Eq.(\ref{schartz C}) also gives $f_{QQ}=0$ and this means that Eq.(\ref{schartz C}) is also not compatible with the coincident gauge . It's a very interesting thing. Since if we fix the coincident gauge, different choices of coordinate systems mean different choices of covariant derivatives that we used in theory. So the  existence of the reparameterization symmetry of radius means that different covariant derivatives represent same physics. And we need to emphasize that it is different from the reparameterization symmetries of diffeomorphism. Where it is the same covariant derivative represented in different coordinate systems.

\subsubsection{FRW universe}

Modified gravity theories should have Friedmann-Robertson-Walker (FRW) solutions for our universe, which is based on the assumption that at large scales the universe is homogeneous and isotropic in space, this was confirmed by observations and becomes the cornerstone of cosmology. For spatially flat FRW universe, from \cite{Jimenez:2019ovq, Lu:2019hra} we know coincident gauge are compatible with the Cartesian coordinate system, which means that in  Cartesian coordinate system, choosing ${\Gamma^{\lambda}}_{\mu\nu} =0$ is a solution of $f(Q)$ theory. But for spatially flat FRW universe in spherical coordinate system and for spatially curved universe, the situation has changed.

If we take the coincident gauge (${\Gamma^{\lambda}}_{\mu \nu} = 0$) and further assume that the coordinate system determined by this gauge is homogeneous and isotropic coordinate system. Then the FRW metric can be written as
\begin{eqnarray} \label{cos metric}
	ds^2 = -N^2(t)dt^2 + a^2(t) \left[   \frac{dr^2}{1-k r^2} +r^2( d\theta^2 + \sin^2\theta d\phi^2 ) \right]
\end{eqnarray}
where $N(t)$ is the lapse function. Taking this metric form into EOMs of metric, i.e. Eq.(\ref{metric f(Q) eq}) and considering the matter is perfect fluid, we can obtain the non-diagonal term
\begin{eqnarray} \label{non-diagonal FRW}
	E_{\theta t} = E_{t\theta} =-\frac{1}{2}f_{QQ}  \frac{\partial Q}{\partial t} \cot\theta =0, \quad E_{\theta r} = E_{r\theta}=-\frac{1}{2}f_{QQ}  \frac{\partial Q}{\partial r} \cot\theta =0
\end{eqnarray}
where $Q = 2\left(  k - 1/r^2- 3a^{'2}/ N^2\right)/a^2$. Thus it is easy to find that if $f_{QQ} \neq 0$, Eq.(\ref{non-diagonal FRW}) has no meaningful solutions. So it faces the same problem as static spherically symmetric spacetime.

\subsection{Method to find solutions}

From precious subsection, we have known that choosing coincident gauge and further assuming spherically symmetric coordinate system in spherically symmetric spacetime or homogeneous and isotropic coordinate system in FRW universe will lead to inconsistent result $f_{QQ}=0$. This does not mean there are no solutions for those case but means the coordinate systems we choose are not compatible with the coincident gauge. So to find the solutions for those two common cases, there are two direct ways
\begin{itemize}
	\item Since the coordinate systems we choose are not compatible with the coincident gauge, we can choose another arbitrary coordinate system $\left\{y^{\mu}\right\}$ in which we think ${\Gamma^{\lambda}}_{\mu\nu} =0$. 
	\item We can also do not choose the coincident gauge and assume the affine connection choose the general form Eq.(\ref{gamma=}). Then look for the suitable four arbitrary functions $y^{\mu}$.
\end{itemize} 
In this paper, we choose the second approach since we prefer the coordinate systems which are constructed by the symmetries of spacetime. In arbitrary coordinate system, the symmetries of spacetime is hidden and we think it may complicate the problems.

If using the second approach to look for solutions straightforward, one will find the process is very complicated just like in $f(T)$ case \cite{Tamanini:2012h good and bad}. Therefore in this paper, to find solutions for $f(Q)$ theory , we make some simplifications and propose a method similar to the one used in Ref. \cite{Krssak:2015oua} for TEGR. The method is as follows. The curvature-free and torsion-free conditions demand there is a coordinate system $\left\{ y^{\mu}\right\}$ in which the connection satisfies ${\Gamma^{\lambda}}_{\mu \nu} (y^{\mu})= 0$, so in arbitrary coordinate system, the affine connection will be Eq.(\ref{gamma=}). It means that the connection with form Eq.(\ref{gamma=}) only relies on the choices of coordinate system and does not relies on the gravity of spacetime. Under specific circumstances, spacetime contains the gravitational effects, such that it is difficult to get a suitable affine connection. Since this affine connection only relies on the choices of coordinate system, we can first look for suitable affine connections for the spacetime which does not contain gravitational effects. This spacetime can be obtained by removing the parameters that contain information of gravity, such as we can set $A(r)=B(r)=0$ for the static spherically symmetric spacetime. Also because affine connection only relies on the choices of coordinate system, we can just take the obtained affine connection as a solution of the original spacetime. Here we just give the procedure which we used in the following contents to look for the suitable affine connections. Combining the examples in the next section, one can find that this process is easy to understand.
\begin{itemize}
	\item Firstly, we write down the metric $g$ of considered question and remove the parameters that contain information of gravity to get a new metric $g_{(r)}$. (From now on, the quantities with a subscript $(r)$ represent they are quantities related to the new metric $g_{(r)}$. )
	\item Secondly, we find out suitable non-metricity tensor $Q_{(r)\alpha\mu\nu}$ which satisfies the curvature-free condition for this 'no gravity' spacetime. (This is the improvement of the procedure used in Ref. \cite{Krssak:2015oua}, where the authors have set the torsion tensor to be zero, so that in FRW universe the spin connection will not satisfy the basic requirement of TEGR, i.e., zero curvature. But one can check that in our method, curvature-free condition is always satisfied.)
	\item Finally, using the Eq.(\ref{Gamma=Gamma+S}) which relates the affine connection with the Levi-Civita connection and non-metricity tensor, we will get the affine connections.
\end{itemize}

The details of our method will be presented in the following examples.

\section{Some Solutions}\label{section solution}

Now we show how to solve the gravitation system in terms of our improved method by two important examples: the spherically symmetric spacetime and the FRW universe.

\subsection{Static Spherically Symmetric Spacetime}

We take the metric form as
\begin{eqnarray} \label{schartz C2}
	ds^2 = -e ^{A(r)} dt^2 + e^{B(r)}dr^2 + C(r)^2( d\theta^2 + \sin^2\theta d\phi^2).
\end{eqnarray}
for static spherically symmetric spacetime, because from this we can explicitly find the symmetries of this spacetime. From previous section, we have known that this ansatz for metric are not compatible with the coincident gauge in the sense of equations of motion.  So we should take other gauge conditions for the affine connection. Now we apply our method as follows. Firstly removing the parameters that contain information of gravity in the metric Eq.(\ref{schwartz metric}), this can be achieved by merely setting $A(r)=0, B(r)=0, C(r)=r$. So that metric Eq.(\ref{schartz C2}) reduces to the Minkowski metric formulated in the spherical coordinate system, 
\begin{eqnarray} \label{close Schwartz metric}
	ds^2 = - dt^2 + dr^2 + r^2( d\theta^2 + \sin^2\theta d\phi^2),
\end{eqnarray}
and we denoted it as $g_{(r)}$. From now on, we consider which affine connection is suitable for the spacetime determined by the metric $g_{(r)}$. Secondly, it is well known that in GR, the curvature tensor ${\mathring{R}_{(r)\mu\nu\rho }}{}^{\sigma}$ which represents the gravity in (Minkowski) spacetime (\ref{close Schwartz metric}) is zero and in STGR theory, it is the non-metricity tensor $Q_{(r)\alpha\mu\nu}$ that represents gravity, so it is a meaningful
assumption by setting $Q_{(r)\alpha\mu\nu}=0$ for this new spacetime determined by $g_{(r)}$. Considering this assumption and the relation between arbitrary affine connection and the Levi-Civita connection Eq.(\ref{Gamma=Gamma+S}), we can get the all non-vanishing components of connection: 
\begin{eqnarray} \label{schwartz gamma}
	{\Gamma^{r}}_{\theta\theta } = -r, \quad {\Gamma^{r}}_{\phi\phi} = -r \sin^2\theta,\nonumber \\ {\Gamma^{\theta}}_{\phi\phi} = -\cos\theta \sin\theta,\quad 
	{\Gamma^{\theta}}_{r\theta} = {\Gamma^{\theta}}_{\theta r}= \frac{1}{r}, \nonumber \\  {\Gamma^{\phi}}_{r\phi} = {\Gamma^{\phi}}_{\phi r}=\frac{1}{r}, \quad {\Gamma^{\phi}}_{\theta\phi} = {\Gamma^{\phi}}_{\phi \theta} =\cot\theta.
\end{eqnarray}
We take this affine connection Eq.(\ref{schwartz gamma}) as the suitable affine connection for static spherically symmetric spacetime in $f(Q)$ theory. 

As an example, we set $C(r)=r$ (just a special case) to see what will happen if we choose the affine connection as Eq.(\ref{schwartz gamma}). Using the metric Eq.(\ref{schartz C2}) in which $C(r)=r$ and connection Eq.(\ref{schwartz gamma}), we can get the equations of motion of $f(Q)$ theory, i.e., $E_{\mu\nu}=\tau_{\mu\nu}$
\begin{eqnarray} \label{eom of f(Q) S}
	&&\tau_{tt}=\frac{e^{A-B}}{2r^2} \{e^{B}r^2f+f_{Q}\left[(e^{B}-1)(2+rA')+(1+e^{B})rB'\right]+2f'_{Q}r(e^{B}-1)\},\nonumber\\
	&&\tau_{rr}=\frac{-1}{2r^2}\{ e^{B}r^2f+f_{Q}\left[(e^{B}-1)(2+rB'+rA')-2rA'\right]+2f'_{Q}r(e^{B}-1) \},\nonumber\\
	&&\tau_{\theta\theta}=-\frac{r}{4e^{B}}\{f_{Q}\left[-4A'-rA^{'2}-2rA''+rA'B'+2e^{B}(A'+B')\right] +2e^{B}rf-2f'_{Q}rA'\}\nonumber\\
	&&\tau_{\phi\phi}=\tau_{\theta\theta}\sin^2\theta,
\end{eqnarray}
where $Q=e^{-B}(1-e^{B})(A'+B')/r$ and $f'_{Q}=f_{QQ}dQ/dr$.
So one can see that using the affine connection obtained here, the field equations are diagonal compared with the case in which $\Gamma^{\lambda}{}_{\mu\nu}=0$ is used, i.e., Eq.(\ref{off-diagonal sphere}). Given a concrete form of $f(Q)$ and boundary conditions we can solve above equations to get the configurations  $A(r)$ and $B(r)$. Such as if  we consider vacuum solutions, this is $\tau_{\mu\nu}=0$, then Eq.(\ref{eom of f(Q) S}) gives us $A'(r)+B'(r)=0$. One can see that in this special case, non-metricity scalar $Q=0$. From the metric's equations of motion Eq.(\ref{new form EOMs}), it means
\begin{eqnarray}
	f_{Q}\mathring{G}_{\mu\nu} -fg_{\mu\nu}/2=0.
\end{eqnarray}
Therefore if  $f_{Q} \neq 0$, such as $f(Q)=Q+\alpha Q^2$, field equations become $\mathring{G}_{\mu\nu}+\Lambda g_{\mu\nu}=0$, which is just the Einstein equations with a cosmological constant term i.e., $\Lambda=-f/(2f_{Q})$. So for vacuum solutions, the affine connection we obtained in this paper leads to same evolution as its in GR. Generally speaking, for non-vacuum case, it may leads to different solutions.

Just like section \ref{action and eoms} said, if EOMs of metric and matter are satisfied, the EOMs of affine connection are satisfied automatically, thus Eq.(\ref{eom of f(Q) S}) contains all information of evolution. So if one does a fully analysis of the Eq.(\ref{eom of f(Q) S}), he does not need consider the equations of affine connection. Since a fully analysis is not the subject of this paper, we also write down the field equations of affine connection, then Eq.(\ref{connection equations}) gives us
\begin{eqnarray}\label{constraint affine}
	\left[(e^{B}-1)(4+rA'+rB')+2rB'\right]f'_{Q}+2(e^{B}-1)rf''_{Q}=0.
\end{eqnarray}
It is easy to see that vacuum solution satisfies this equation automatically. For other case, this equation may be a constraint. The existence of this equations is easy to understand. If the EOMs of matter fields are satisfied, the energy-momentum conservation is satisfied, $\mathring{\nabla}_{\mu}\tau^{\mu\nu}=0$. But the gravitational part of field equations of metric $E^{\mu\nu}$ maybe not satisfy the equation $\mathring{\nabla}_{\mu}E^{\mu\nu}=0$ and this will lead constraints like Eq.(\ref{constraint affine}). (In GR, because the gravitational part of action only depends on metric and also is diffeomorphism invariant, $\mathring{\nabla}_{\mu}\mathring{G}^{\mu\nu}=0$ is an identity. But here the gravitational part of action also depends on affine connection.) It seems that Eq.(\ref{connection equations}) gives us strong constraint of models. But until now, we set $C(r)=r$ for simplicity. If we keep $C(r)$, the field equations of metric are also diagonal and the equations of affine connection becomes
\begin{eqnarray}\label{spherical connection C equation}
	\left[(e^{B}r^2-C^2)(4+rA'+rB')+4C^2-4rCC'+2C^2rB'\right]f'_{Q} +2(e^{B}r^2-C^2)rf''_{Q}=0
\end{eqnarray}
Thus this equation will become an equation that determine the evolution of function $C(r)$. Therefore for general cases, we may need keep $C(r)$ to satisfy the EOMs of affine connection.

Here we count the number of independent equations and free functions to see if this will lead to an overdetermined system of equations. For spherically symmetric case, there are three (metric) plus one (affine connection) equations. But there is also an identity which can be derived from diffeomorphism invariance (just like Bianchi identity in GR), so the number of independent equations is three. Because there are three free functions $A(r)$, $B(r)$, $C(r)$, we think this will not lead to an overdetermined system.

Just as section\ref{action and eoms} said, there is a coordinate system in which $\Gamma=0$ and it is easy to see that Eq.(\ref{schwartz gamma}) is just the Levi-Civita connection of Minkowski spacetime in a spherical coordinate system, so that in Cartesian coordinate system it becomes zero. The coordinate transformation is
\begin{eqnarray}
	y^{0} = t, \quad y^{1} = r \sin\theta \cos\phi, \quad y^{2} = r \sin\theta \sin\phi, \quad y^{3} = r \cos\theta,
\end{eqnarray}
and the coordinate system $\left\{y^{\mu}\right\}$ is the one in which $\Gamma=0$. In this coordinate system, metric can be rewritten as
\begin{eqnarray}
	ds^2 = - e^{A} dt^2 +\frac{C^2}{r^2}\left[d(y^{1})^{2} +d(y^{2})^2 + d(y^{3})^2\right] +\frac{e^{B}r^2-C^2}{r^4}(y^{i}y^{j} dy^{i} dy^{j}),
\end{eqnarray}
where $i=1,2,3$ and one can see it is more complicated than Eq.(\ref{schwartz metric}) and difficult to find out the symmetries of the spacetime, though it is the same tensor as Eq.(\ref{schwartz metric}).

\subsection{FRW Universe}

\subsubsection{Spatially flat FRW universe $k=0$}
The spatially flat FRW solution have the following metric form in the spherical coordinate system,
\begin{eqnarray} \label{cos metric1}
	ds^2 = -N^2(t)dt^2 + a^2(t) \left[   dr^2 +r^2( d\theta^2 + \sin^2\theta d\phi^2 ) \right]~,
\end{eqnarray}
and in Cartesian coordinate system the metric form becomes
\begin{eqnarray} \label{k=0 metric}
	ds^2 = -N^2(t)dt^2 + a^2(t)(dx^2+dy^2+dz^2).
\end{eqnarray}
Follow the above procedure, we can find suitable affine connection in this case. Firstly removing the parameters that contain information of gravity by setting $a(t)=1, N(t)=1$, so that Eq.(\ref{k=0 metric}) becomes the Minkowski metric in Cartesian coordinate. Then just as Minkowski metric in spherical coordinate case, we also choose $Q_{(r)\alpha\mu\nu}=0$ for this new spacetime. Finally using Eq.(\ref{Gamma=Gamma+S}), we obtain that all components of connection are zero. This is to say that Cartesian coordinate system is just the one in which connection vanishes and the coincident gauge causes no inconsistency. In fact this has been done in many cosmological applications \cite{Jimenez:2019ovq, Lu:2019hra}.

For $k=0$ spherical coordinate system case, the condition $a(t)=1$ makes Eq.(\ref{cos metric1}) the same as Eq.(\ref{close Schwartz metric}), so that the suitable affine connection also is Eq.(\ref{schwartz gamma}). This result have been obtained in \cite{Runkla:2018xrv}

Up to now, one can find that all above examples share the same property that if we remove the parameters that contain information of gravity, the considered metrics will reduce to the Minkowski metric formulated in various coordinate systems. Therefore we can always choose $Q_{(r)\alpha\mu\nu}=0$ to represent that the Minkowski spacetime does not contain gravity. Then the obtained connections are just the Levi-Civita connections of Minkowski metric in different coordinate systems, so that the curvature-free condition is satisfied automatically. But in spatially curved FRW universe, the situation is different. 

\subsubsection{$k=\pm1$ curved space case} 

Removing the parameters that contain information of gravity by setting $a(t)=1, N(t)=1$ in $k = \pm 1$ curved space case, we get the metric $g_{(r)}$
\begin{eqnarray} \label{k=1 close}
	ds^2 = -dt^2 + \left[   \frac{dr^2}{1-k r^2} +r^2( d\theta^2 + \sin^2\theta d\phi^2 ) \right].
\end{eqnarray}
Of course Riemann tensor from the Levi-Civita connection of this metric Eq.(\ref{k=1 close}) does not vanish. This is to say even we choose $a(t)=1$ in GR, the spacetime is still curved and the gravity still exists. Therefore considering the curvature relation Eq.(\ref{curvature relation}) in STGR theory, if we still take $Q_{(r)\alpha\mu\nu}=0$ for this spacetime, the obtained affine connections will not satisfy the curvature-free condition. So that we should consider the question that how to choose a suitable non-metricity tensor for this case.

To get a non-metricity tensor which is compatible with vanishing curvature and torsion, a direct idea is to solve the Eq.(\ref{curvature relation}). But $Q_{(r)\alpha\mu\nu}$ owns too many components so that it is difficult to solve it, therefore we need add some constraints on it. Here we give the constraint with following considerations. If we denote $\eta^\mu$ as the Killing vectors of metric $g_{(r)}$ i.e. Eq.(\ref{k=1 close}), the constraint we require is that for all Killing vectors $\eta^\mu$, the Lie derivatives of $Q_{(r)\alpha\mu\nu}$ are zero, that is 
\begin{eqnarray} \label{Q's condition}
	\mathcal{L}_{\eta} Q_{(r)\alpha\mu\nu} = 0.
\end{eqnarray}
The argument why we choose this condition is as follows. The cosmological principle says that the universe is spatially homogeneous and isotropic. From the monograph \cite{Weinberg cosmology}, we can see that this assumption means
\begin{itemize}
	\item the three dimension subspace of constant cosmic time is maximally symmetric subspace, and it owns six Killing vectors, that we denote them as $\zeta^{\mu}$,
	\item all cosmic tensor such as the metric $g_{\mu\nu}$, the energy momentum tensor $T_{\mu\nu}$, and so on are form invariant with respect to all the isometries of these subspaces, which means for all cosmic tensor $W$, $\mathcal{L}_{\zeta} W =0$.
\end{itemize}
So now in $f(Q)$ theory, it is a natural idea to consider a form invariant non-metricity tensor. For the spacetime i.e. Eq.(\ref{cos metric}), a meaningful form of  non-metricity tensor should satisfy $	\mathcal{L}_{\zeta} Q_{\alpha\mu\nu} = 0$. We extend this idea and require that for any spacetime, suitable non-metricity tensors should satisfy the condition that the Lie derivatives of them vanish for all Killing vectors of this spacetime.

The form of non-metricity tensor which satisfies the cosmological symmetries has already been derived in \cite{Minkevich:1998cv}. But we have not found a proof, so we give a proof  in \ref{proof of Q form}. A better form was given in \cite{Iosifidis:2020gth} in which the non-metricity tensor which satisfies $\mathcal{L}_{\zeta} Q_{(r)\alpha\mu\nu} =0$ for all six spatial Killing vectors $\zeta^{\mu}$ was written as
\begin{eqnarray} \label{Q' form}
	Q_{(r)\alpha\mu\nu} = A(t) U_{\alpha} h_{(r)\mu\nu} +B(t) h_{(r)\alpha(\mu} U_{\nu)} +C(t) U_{\alpha}U_{\mu}U_{\nu},\nonumber\\
\end{eqnarray}
where $A(t), B(t), C(t)$ are three arbitrary functions, $U_{\alpha} = (dt)_{\alpha}$, $h_{(r)\mu\nu}=g_{(r)\mu\nu}+U_{\mu}U_{\nu}$ is the induced metric. Then using Eq.(\ref{Gamma=Gamma+S}) we can get all non-vanishing components of connection,
\begin{eqnarray} \label{k=1 gamma}
	{\Gamma^{t}}_{tt} = \mathcal{K}_{1}, \quad {\Gamma^{t}}_{rr} = \frac{\mathcal{K}_{2}}{\chi^{2}}, \quad {\Gamma^{t}}_{\theta\theta}=\mathcal{K}_{2}r^2,\nonumber\\ {\Gamma^{t}}_{\phi\phi}=\mathcal{K}_{2} r^{2} \sin^{2}\theta,\quad {\Gamma^{r}}_{rr}=\frac{kr}{\chi^2}, \nonumber\\
	{\Gamma^{\theta}}_{\theta r}={\Gamma^{\theta}}_{r\theta}={\Gamma^{\phi}}_{\phi r}={\Gamma^{\phi}}_{r\phi}=\frac{1}{r}, \nonumber\\ {\Gamma^{\phi}}_{\phi\theta}={\Gamma^{\phi}}_{\theta\phi}=\cot\theta, \quad {\Gamma^{\theta}}_{\phi\phi} = -\sin\theta \cos\theta, \nonumber\\
	{\Gamma^{r}}_{\theta\theta}= -r\chi^{2}, \quad {\Gamma^{r}}_{\phi\phi}=-r\chi^{2}\sin^{2}\theta,\nonumber\\
	{\Gamma^{r}}_{tr}={\Gamma^{r}}_{rt}={\Gamma^{\theta}}_{t\theta}={\Gamma^{\theta}}_{\theta t}={\Gamma^{\phi}}_{t\phi}={\Gamma^{\phi}}_{\phi t}=\mathcal{K}_{3},
\end{eqnarray}
where $\chi^{2}=1-kr^{2}$, $\mathcal{K}_{1}=-C/2, \mathcal{K}_{2}=(A-B)/2, \mathcal{K}_{3}=A/2$. Similar forms can be found in \cite{Minkevich:1998cv, Hohmann:2019fvf}. Meanwhile this connection should satisfy the curvature-free condition and this gives the constraints about three free parameters $\mathcal{K}_{1}(t), \mathcal{K}_{2}(t), \mathcal{K}_{3}(t)$
\begin{eqnarray}
	\mathcal{K}_{3} (\mathcal{K}_{1} - \mathcal{K}_3) - \dot{\mathcal{K}}_{3}  &=0 , \nonumber\\
	\mathcal{K}_{2}(\mathcal{K}_{1} - \mathcal{K}_{3}) + \dot{\mathcal{K}}_{2} &= 0 ,\nonumber\\
	k +\mathcal{K}_{2}\mathcal{K}_{3} &=0.
\end{eqnarray}
Until now, we have only used the spatial symmetries and for $a(t)=1$ case, there are more symmetries. For $k=0$ case, we know it is just the Minkowski spacetime which is a maximally symmetric spacetime, then the condition Eq.(\ref{Q's condition}) requires $Q_{(r)\alpha\mu\nu} = 0$  (the proof has been given in \ref{proof k=0, Q=0}). Therefore for this case, it returns to what we have done above. For $k=\pm1$ case, there is only one additional Killing vector $(\partial / \partial {t})^{\mu}$ and the condition Eq.(\ref{Q's condition}) for this Killing vector requires that $\mathcal{K}_{1}, \mathcal{K}_{2}, \mathcal{K}_{3}$ are all constants. So that for this case we finally obtain
\begin{eqnarray}
	\mathcal{K}_1 = \mathcal{K}_{3} = -\frac{k}{\mathcal{K}_{2}}.
\end{eqnarray}
For $k=\pm1$ case, the proof that the metric $g_{(r)}$ (Eq.(\ref{k=1 close})) only has seven Killing vectors can be found in \ref{proof no other Killing vector}.

As an example, we can use this affine connection to get the EOMs of $f(Q)$ for FRW universe. For simplicity, we first set $N(t)=1$, then the field equations of metric are
\begin{eqnarray}
	\frac{1}{2}\bigg( f+3H(4H-3\mathcal{K}_{1})f_{Q}+3\mathcal{K}_{1}f'_{Q} -\frac{3\mathcal{K}_{2}(Hf_{Q}+f'_{Q})}{a^{2}} \bigg) =\rho, \nonumber\\
	\frac{1}{2}\bigg(  f +H(8H-9\mathcal{K}_{1})f_{Q} +(4H-3\mathcal{K}_{1})f'_{Q} -\mathcal{K}_{2}\frac{(3H-4\mathcal{K}_{1})f_{Q} + f'_{Q}}{a^{2}} + 4f_{Q}\frac{a''}{a} \bigg)=-P,
\end{eqnarray}
where $a'(t)=\partial a/\partial t, f'_{Q}=\partial f_{Q}/\partial{t}$ and $Q=-6H^{2}+9H\mathcal{K}_{1}+ 3\mathcal{K}_{2}(H-2\mathcal{K}_{1} )/a^{2}$. Compared with the case using the coincident gauge, here metric field equation $E^{\mu\nu}$ is diagonal. We also can write down the field equation for affine connection, which is
\begin{eqnarray}
	(k+3\mathcal{K}_{3}^2a^2)a'f'_{Q}+	(k+\mathcal{K}_{3}^2a^2)af''_{Q}=0.
\end{eqnarray}
Just like in spherically symmetric spacetime, it seems that this equation gives us a constraint. Therefore if the models we considered do not satisfy this equations, we can keep the function $N(t)$ to make this equation an equation that determines function $N(t)$. If we keep function $N(t)$, It is easy to find that the metric field equations is also diagonal and the equations of affine connection becomes
\begin{eqnarray}
	(3\mathcal{K}_{3}^{2}a^2N+kN^3-\mathcal{K}_{3}^{2}a^3N'/a'+kaN^2N'/a')a'f'_{Q}
	+(\mathcal{K}_{3}^2a^2N+kN^3)af''_{Q}=0
\end{eqnarray}
which will be the equation that determines the evolution of function $N(t)$. Also like in spherically symmetric spacetime case, there are two (metric) plus one (affine connection) plus one (matter field) minus one (identity comes from diffeomorphism invariance) equals three independent equations and $a(t)$, $N(t)$, $\rho(t)$, $P(t)$ four functions. Then given the equation of state $P=P(\rho)$, we can solve this system.

\section{Conclusion} \label{summery}

In $f(Q)$ theory, the curvature-free and torsion-free conditions always allow us to choose the so-called 'coincident gauge' in which affine connection vanishes. In this paper, we pointed out that taking the coincident gauge in  $f(Q)$ theory sometimes makes us unable to choose the coordinate systems based on symmetries. We used two examples to illustrate this, one is the static spherically symmetric spacetime and the other is FRW universe. For both cases, we found that if at the beginning $f(Q)$ is not a linear function of $Q$, such as $f(Q)=Q^2$, the coincident gauge and the coordinate system based on the symmetries of spacetime are not compatible. Such as in FRW universe case, ${\Gamma^{\lambda}}_{\mu\nu}=0$ and the metric expanded in homogeneous and isotropic coordinate system is not a solution of $f(Q)$ theory.

There are two approaches to solve this problem. One is that since the coordinate systems based on symmetry are not compatible with the coincident gauge, we choose other coordinate systems. The other is that we do not choose the coincident gauge and assume the affine connection takes the general form. In this paper, we choose the second approach and proposed a improved method to search for suitable affine connections in $f(Q)$ theory given a metric ansatz.  We applied this method to the static spherically symmetric spacetime and FRW universe. We found that in  static spherically symmetric spacetime, the obtained affine connection is just the Levi-Civita connection of Minkowski spacetime formulated in spherical coordinate system and we also got the coordinate system which is compatible with coincident gauge. In FRW universe case, after requiring the non-metricity tensor satisfies all the symmetries of the spacetime which scale factor is set to one, we also obtained the suitable affine connections. Using the method in this paper, we can not only find the solutions for $f(Q)$ theory, but also for all other modified STGR theories. We also should point out that there may be other solutions for $f(Q)$ theory beyond the ones found using the method in this paper, since in the process of getting solutions we have made some simplifications. But because it is too difficult to find all solutions for $f(Q)$ theory, so we think the solutions found in this paper are important. This method and its various applications deserve further studies.

\begin{acknowledgements}
	The author would like to thank Mingzhe Li and Haomin Rao for useful discussions. This work is supported by NSFC under Grant Nos. 12075231, 11653002, 12047502 and 11947301.
\end{acknowledgements}

\appendix

\section{Nice form of EOMs of $f(Q)$}\label{prof of nice formula}

In this appendix, we rewrite metric's EOMs of $f(Q)$ theory i.e. Eq.(\ref{metric f(Q) eq}) as a new form i.e. Eq.(\ref{new form EOMs}). Since $\mathring{\nabla}$ is the covariant derivative associated with Levi-Civita connection, we have
\begin{eqnarray}
	\mathring{\nabla}_{\lambda}\sqrt{-g} = \partial_{\lambda} \sqrt{-g} - {\mathring{\Gamma}^{\sigma}}{}_{\sigma\lambda}\sqrt{-g}=0.
\end{eqnarray}
The curvature-free and torsion-free $\nabla$ satisfies
\begin{eqnarray}
	&&\nabla_{\lambda}\sqrt{-g} = \partial_{\lambda} \sqrt{-g} - {\Gamma^{\sigma}}_{\sigma\lambda}\sqrt{-g} ,\\
	&&\nabla_{\lambda}\sqrt{-g} - \mathring{\nabla}_{\lambda}\sqrt{-g}  = \nabla_{\lambda}\sqrt{-g} = -( {\Gamma^{\sigma}}_{\sigma\lambda}- {\mathring{\Gamma}^{\sigma}}{}_{\sigma\lambda})\sqrt{-g} = \frac{1}{2} \sqrt{-g}{Q_{\lambda\sigma}}^{\sigma},
\end{eqnarray}
so that the first term of Eq.(\ref{metric f(Q) eq}) can be rewritten as
\begin{eqnarray}
	\frac{2}{\sqrt{-g}} \nabla_{\lambda} (\sqrt{-g} f_{Q} {P^{\lambda}}_{\mu\nu})=f_{Q} {Q_{\lambda \sigma}}^{\sigma} {P^{\lambda}}_{\mu\nu} +2(f_{Q}\nabla_{\lambda}{P^{\lambda}}_{\mu\nu} +f_{QQ} \partial_{\lambda}Q {P^{\lambda}}_{\mu\nu} ).
\end{eqnarray}
From the curvature relation i.e. Eq.(\ref{curvature relation}), we can get the relations between curvature $\mathring{R}_{\mu\nu}$ and non-metricity $Q_{\alpha\mu\nu}$
\begin{eqnarray}
	\mathring{R}_{\mu\nu} &=& \nabla_{\mu}{N^{\sigma}}_{\sigma\nu} - \nabla_{\sigma}{N^{\sigma}}_{\mu\nu} + {N^{\lambda}}_{\mu\nu}{N^{\sigma}}_{\sigma\lambda} - {N^{\lambda}}_{\sigma\nu}{N^{\sigma}}_{\mu\lambda},\nonumber\\
	\mathring{R} &=& g^{\mu\nu}	\mathring{R}_{\mu\nu},
\end{eqnarray}
where ${N^{\lambda}}_{\mu\nu}= -\frac{1}{2}({Q_{\mu\nu}}^{\lambda} + {Q_{\nu\mu}}^{\lambda} -{Q^{\lambda}}_{\mu\nu})$.
Then after some calculations, we can get
\begin{eqnarray}
	\mathring{G}_{\mu\nu}  &= &\mathring{R}_{\mu\nu} -\frac{1}{2} \mathring{R}g_{\mu\nu} \nonumber\\
	&=&2 \nabla_{\lambda} {P^{\lambda}}_{\mu\nu} -\frac{1}{2}Qg_{\mu\nu} + (P_{\rho\mu\nu}{Q^{\rho\sigma}}_{\sigma} +P_{\nu\rho\sigma}{Q_{\mu}}^{\rho\sigma} -2P_{\rho\sigma\mu}{Q^{\rho\sigma}}_{\nu}       ),\nonumber
\end{eqnarray}
Finally, we obtain the new form of EOMs of $f(Q)$ theory:
\begin{eqnarray} 
	f_{Q} \mathring{G}_{\mu\nu} +\frac{1}{2} g_{\mu\nu} (f_{Q}Q - f) +2f_{QQ}(\partial_{\lambda}Q) {P^{\lambda}}_{\mu\nu} = \tau_{\mu\nu},
\end{eqnarray}

\section{The form of non-metricity tensor in FRW universe} \label{proof of Q form}
In this appendix, we give a proof that the form of  $Q_{\alpha\mu\nu}$ which satisfies the cosmological symmetries $\mathcal{L}_{\zeta} Q_{\alpha\mu\nu} =0$ is
\begin{eqnarray} \label{Q' form 2}
	Q_{\alpha\mu\nu} = A(t) U_{\alpha} h_{\mu\nu} +B(t) h_{\alpha(\mu} U_{\nu)} +C(t) U_{\alpha}U_{\mu}U_{\nu},
\end{eqnarray}
where $h_{\mu\nu}$ is induced metric of  FRW metric (\ref{cos metric}) , $h_{\mu\nu}=g_{\mu\nu}+U_{\mu}U_{\nu}$.
Since the three dimensional space is maximally symmetric in FRW universe, so that we can use the trick in the monograph \cite{Weinberg cosmology} to handle it and one can also find that this trick has been used to get the form of torsion tensor in reference \cite{Cosmological principle and torsion}.

For non-metricity tensor $Q_{\alpha\mu\nu}$, we can decompose it into spatial scalar $Q_{000}$, spatial vectors $ Q_{i00}, Q_{0i0}, Q_{00i}$, spatial rank-2 tensors $ Q_{ij0}=Q_{i0j}, Q_{0ij}$, and spatial rank-3 tensor $Q_{ijk}$. Where we use $\alpha, \mu, \nu,...=0,1,2,3$ as spacetime indices and $i,j,k,m,n...=1,2,3$ as space indices. Therefore the condition that for all six spatial Killing vectors $\zeta^{\mu}$, $\mathcal{L}_{\zeta} Q_{\alpha\mu\nu} =0$, now is equivalent to that the Lie derivatives of all components of $Q_{\alpha\mu\nu}$  are zero. 

First we calculate $\mathcal{L}_{\zeta} Q_{ijk} =0$.
\begin{eqnarray} \label{space Q condition}
	\mathcal{L}_{\zeta} Q_{ijk}& =&\zeta^{m} \mathring{D}_{m}Q_{ijk} +\mathring{D}_{n}\zeta_{m}
	( \delta^{n}_{i} {Q^{m}}_{jk}+\delta^{n}_{j}{{Q_{i}}^{m}}_{k}+\delta^{n}_{k}{Q_{ij}}^{m} )=0,
\end{eqnarray}
where $\mathring{D}$ represents the spatial covariant derivative associated with the induced metric $h_{\mu\nu}$ and the corresponding connection is the Levi-Civita connection. We use $h_{\mu\nu}$ to raise and lower index. Because the space is maximally symmetric, so that at any point for those Killing vectors which represent rotations, we can always make them satisfy the requirements that $\zeta^{i}=0$ and $\mathring{D}_{n}\zeta_{m}$ is anti-symmetric. Therefore the condition Eq.(\ref{space Q condition}) is equivalent to
\begin{eqnarray}\label{space Q condition 2}
	\delta^{n}_{i} {Q^{m}}_{jk}+\delta^{n}_{j}{{Q_{i}}^{m}}_{k}+\delta^{n}_{k}{Q_{ij}}^{m} = \delta^{m}_{i} {Q^{n}}_{jk}+\delta^{m}_{j}{{Q_{i}}^{n}}_{k}+\delta^{m}_{k}{Q_{ij}}^{n} .
\end{eqnarray}
One might think that we have only used three of the six Killing vectors, but it is not the case. Certainly there are six Killing vectors in FRW universe, three for rotations and three for translations. The presence of three Killing vectors representing rotations at one point means that the space is isotropic about this point and three for translations means the space is homogeneous about the point. Meanwhile, any space that is isotropic about every point is also homogeneous. So in the progress of getting Eq.(\ref{space Q condition 2}), we have used all conditions. The contraction of $n,i$ gives
\begin{eqnarray}\label{contract0}
	2Q_{mjk} +Q_{jmk} + Q_{kjm} = h_{mj}{{Q_{i}}^{i}}_{k} + h_{mk} {Q_{ij}}^{i},
\end{eqnarray}
the contraction of $m,j$ gives 
\begin{eqnarray}\label{contract1}
	{Q_{ki}}^{i} = {Q_{ik}}^{i},
\end{eqnarray}
and the contraction of $j,k$ gives
\begin{eqnarray}\label{contract2}
	2{Q_{ki}}^{i}=0.
\end{eqnarray}
Then bringing Eq.(\ref{contract1}) and Eq.(\ref{contract2}) into Eq.(\ref{contract0}) , we can get
\begin{eqnarray}
	2Q_{mjk} +Q_{jmk} + Q_{kjm} = 0,
\end{eqnarray}
and exchanging $m$ and $k$
\begin{eqnarray}
	2Q_{kjm} +Q_{jkm} + Q_{mjk} = 0,
\end{eqnarray}
finally we have
\begin{eqnarray}
	Q_{mjk}=Q_{kjm}.
\end{eqnarray}
So it means that $Q_{ijk}$ is totally symmetric tensor. But considering Eq.(\ref{contract0}), it must be zero.

For scalars, vectors, rank-2 tensors, using the same tricks as above, monograph\cite{Weinberg cosmology} has already given the answers.  Finally we can write down all non-vanishing components of $Q_{\alpha\mu\nu}$,
\begin{eqnarray}
	Q_{000}=C(t), \quad Q_{i00}=Q_{0i0}=Q_{00i} =0, \quad Q_{ij0}=Q_{i0j}=\frac{B(t)}{2}h_{ij}, \quad Q_{0ij} =A(t)h_{ij}.
\end{eqnarray}
This is just the result we want.

\section{The proof that in k=0 case, $Q_{(r)\alpha\mu\nu}=0$} \label{proof k=0, Q=0}
In $k=0$ case, the spacetime determined by metric Eq.(\ref{k=1 close}) is not only the space is maximally symmetric but also the whole spacetime. So we can also use the method used in \ref{proof of Q form} for the total non-metricity tensor $Q_{(r)\alpha\mu\nu}$.  We obtain the same relation just as Eq.(\ref{space Q condition 2})
\begin{eqnarray}
	\delta^{\lambda}_{\alpha} {{Q_{(r)}}^{\sigma}}_{\mu\nu} +\delta^{\lambda}_{\mu} {{Q_{(r)\alpha}}^{\sigma}}_{\nu} +\delta^{\lambda}_{\nu} {Q_{(r)\alpha\mu}}^{\sigma} = 	\delta^{\sigma}_{\alpha} {{Q_{(r)}}^{\lambda}}_{\mu\nu} +\delta^{\sigma}_{\mu} {{Q_{(r)\alpha}}^{\lambda}}_{\nu} +\delta^{\sigma}_{\nu} {Q_{(r)\alpha\mu}}^{\lambda} .
\end{eqnarray}
Then using the same procedure given in Appendix.\ref{proof of Q form}, we have $Q_{(r)\alpha\mu\nu}=0$.

\section{There are only seven Killing vectors in $k=\pm 1$ case.} \label{proof no other Killing vector}

In this appendix, we give the proof that metric Eq.(\ref{k=1 close}) only has 7 Killing vectors in $k=\pm 1$ case. And only in this appendix, we omit the ring on the head of  $\mathring{\nabla}$, this is to say, all quantities are associated with the Levi-Civita connection in the following content.

The spacetime which metric is Eq.(\ref{k=1 close}) is not maximally symmetric spacetime, since it does not meet the condition $R_{\mu\nu\rho\sigma} = K(g_{\mu\rho}g_{\nu\sigma} - g_{\mu\sigma}g_{\nu\rho})$ which a maximally symmetric spacetime should satisfy, where $K$ is a constant. So that the number of Killing vectors for this case is less than 10. From monograph \cite{Weinberg cosmology} we can see that Killing vectors are totally determined by $\zeta_{\mu}(X)$ and $\nabla_{\mu}\zeta_{\nu}(X)$ at any point $X$ of a spacetime. Therefore the number of Killing vectors of a spacetime is determined by the number of independent dimensions spanned by  $\zeta_{\mu}(X)$ and $\nabla_{\mu}\zeta_{\nu}(X)$. So that for $N$ dimension maximally symmetric spacetime, the number of Killing vectors is $N + (N-1)N/2$, where $N$ comes from $\zeta_{\mu}(X)$ and $ (N-1)N/2$ comes from anti-symmetric tensor $\nabla_{\mu}\zeta_{\nu}(X)$. For non-maximally symmetric spacetime, we should look for if there are relations between $\zeta_{\mu}(X)$ and $\nabla_{\mu}\zeta_{\nu}(X)$.

Using the definition of Killing vector and cyclic relation of curvature,
\begin{eqnarray}
	\nabla_{\mu}\nabla_{\rho} \zeta_{\sigma} - \nabla_{\rho}\nabla_{\mu} \zeta_{\sigma} ={R_{\mu\rho\sigma}}^{\lambda} \zeta_{\lambda} \nonumber,\\
	{R_{[  \mu\nu\rho]}}^{\lambda}=0 \nonumber,\\
	\nabla_{\mu} \zeta_{\nu} +\nabla_{\nu}\zeta_{\mu}=0,
\end{eqnarray}
we have
\begin{eqnarray}
	\nabla_{\sigma}\nabla_{\rho} \zeta_{\mu} = {R_{\mu\rho\sigma}}^{\lambda}\zeta_{\lambda},
\end{eqnarray}
then
\begin{eqnarray}
	&\nabla_{\nu}&\nabla_{\sigma}( \nabla_{\rho}\zeta_{\mu} ) - \nabla_{\sigma}\nabla_{\nu}( \nabla_{\rho}\zeta_{\mu} ) \nonumber\\&=& {R_{\nu\sigma\mu}}^{\lambda} \nabla_{\rho}\zeta_{\lambda} + {R_{\nu\sigma\rho}}^{\lambda} \nabla_{\lambda}\zeta_{\mu} \nonumber\\
	&=& \nabla_{\nu}({ R_{\mu\rho\sigma}}^{\lambda}\zeta_{\lambda }    )  -\nabla_{\sigma} ({R_{\mu\rho\nu}}^{\lambda}\zeta_{\lambda }  ).
\end{eqnarray}
So we find the final equation that related $\zeta_{\mu}(X)$ and $\nabla_{\mu}\zeta_{\nu}(X)$
\begin{eqnarray} \label{relation zeta and nabla zeta}
	({ R_{\nu\sigma\mu}}^{\lambda} \delta^{\kappa}_{\rho} + {R_{\nu\sigma\rho}}^{\lambda} \delta^{\kappa}_{\mu}- {R_{\mu\rho\sigma}}^{\lambda} \delta^{\kappa}_{\nu} +{R_{\mu\rho\nu}}^{\lambda} \delta^{\kappa}_{\sigma} ) \nabla_{\kappa}\zeta_{\lambda} 
	= ( \nabla_{\nu}{R_{\mu\rho\sigma}}^{\lambda} -\nabla_{\sigma}{R_{\mu\rho\nu}}^{\lambda} )  \zeta_{\lambda}.
\end{eqnarray}
Then in $k=0$ the maximally symmetric spacetime, Eq.(\ref{relation zeta and nabla zeta}) gives $0=0$. But in $k=\pm 1$ case,  it gives three relations, $\nabla_{0}\zeta_{1}=\nabla_{0}\zeta_{2}=\nabla_{0}\zeta_{3}=0$. Therefore in $k=\pm 1$ case the final number of Killing vectors is $10-3=7$, six of them are spatial Killing vectors $\zeta^{\mu}$ and the remaining one is $\partial/\partial t$.

\providecommand{\href}[2]{#2}\begingroup\raggedright\endgroup

\end{document}